August 26, 2011

# Gravity Sources in a Quantum Milne Universe

Geoffrey F. Chew

*Theoretical Physics Group*
*Physics Division*
*Lawrence Berkeley National Laboratory*
*Berkeley, California 94720, U.S.A.*

**Summary**

A quantum-cosmology-suited *sliced* Milne spacetime, located inside a 'big-bang' forward lightcone, comprises the interiors of a sequence of 4-dimensional slices whose invariant 'age' width is at Planck scale. The age of any lightcone-interior point is its Minkowski distance from the lightcone vertex, but *excluded* from the spacetime of a 'quantum Milne universe' (*QMU*) are 3-dimensional *slice-separating*, ray-carrying hyperboloids.

Each (fixed-age) slice boundary houses a Gelfand-Naimark *unitarily* Lorentz-transformable cosmological-Fock-space ray—a sum of tensor products of complex normed fiber-bundle 'source' functions. Each bundle is a unit 3-sphere (fiber) over an invariantly metricized 3-hyperboloid (base space). The evolving *QMU* comprises a slice-interior sequence (each interior a 4-manifold) and a ray sequence housed by the slice-separating 3-manifolds. Throughout each slice interior, the preceding ray specifies classical-field *reality*. Action, of elsewhere-prescribed slice-traversing Feynman paths, determines any ray after the first from its predecessor.

*QMU*s related by a *global* 7-parameter Lie-group symmetry transformation, at fixed age, are equivalent--the *same* quantum universe. This 'cosmological relativity' implies vanishing of *total* universe angular momentum (Mach's principle) as well as of total universe momentum. Milne's cosmological principle accompanies vanishing of total-universe energy. (Seven 'Noether-conserved' quantities *separately* aggregate to zero).

Any *QMU* ray specifies 'immediate-future' local reality through continua of self-adjoint-operator *expectations* that prescribe retarded real (classical) fields over the immediately-following (4-dimensional) slice interior. Exemplifying *QMU* reality is a divergenceless Lorentz symmetric-tensor conserved energy-momentum current density--the Dalembertian divided by *G* of a retarded gravitational potential. Conserved electric-charge current density is the Dalembertian of a retarded electromagnetic 4-vector potential.

**Introduction**

'Relativistic' quantum theory, thus far, has not only lacked Dirac's self-adjoint Hilbert-space operators whose spectra (he proposed) define 'reality' [1] but has failed to recognize a global *arrowed* time. Further, throughout quantum theory's first century a 'Copenhagen' *probabilistic* Hilbert-space interpretation has perplexingly associated reality's meaning to that of 'measurement by conscious observer'. Correction of the foregoing flaws is here achieved by identifying the 'quantum' content of Milne's universe with 'retarded' evolution-propelling *sources* of gravity. The acronym *QMU* will here be employed for 'quantum Milne universe'.

'Universe age'— *QMU*'s anchor-- enjoys a purely-classical scale-setting (see Appendix) status as a symmetry (Lie) group *invariant* that parallels the global (*single*) time of nonrelativistic Dirac quantum theory--*not* associating to any operator's spectrum. Admitted by a heretofore unutilized *fiber-bundle* Fock space, that *unitarily* represents a 7-parameter symmetry group (*7-SG*), are self-adjoint *source-time* operators whose canonically-conjugate (Dirac-sense) partners are the conserved *energies* of individual gravity sources. Six other conserved local *7-SG* generators represent source momentum and angular momentum.

Meaning for continuous-spacetime 'reality' is *QMU*-achieved through classical fields that are *expectations* of self-adjoint field operators with respect to a source-bundle Fock-space ray at an age 'slightly' earlier than the field-point age. Exemplifying the classical fields that define 'settled local reality' is a divergenceless second-rank symmetric Lorentz-tensor--*conserved energy-momentum current density*. This tensor field is proportional, through Newton's gravitational constant, to the gravitational potential's Dalembertian.

The reader is cautioned against confusing the Hilbert-space term 'expectation'-- a real number prescribed by the pairing of a ray and some self-adjoint operator--with this term's ordinary-language significance. *Absent* from our meaning for 'quantum-cosmological expectation' is any probabilistic or consciousness aspect.

Gelfand-Naimark (*GN*) *unitary* Hilbert-space representation of the Lorentz-group [2] fits with Milne's meaning for (classical, continuous) 'spacetime', [3] rather than with Einstein's meaning (either 1905 or 1915), and defines a source-interpretable bundle Fock space at each member of a sequence of exceptional universe ages. Each bundle associates a 3-sphere to every location in an invariantly metricized 3-hyperboloid. Throughout '*SS*'--a Lorentz-group-based *sliced* spacetime that occupies 'much of' but not 'all' the interior of a forward lightcone--the (positive) Lorentz-invariant (continuous) 'age' of any spacetime point is its Minkowski distance from the lightcone's vertex. Each source bundle belongs to *one* member of a discrete sequence of *exceptional* ages.

Milne *age*--foundational to *QMU*--revived a Newtonian idea which a century ago, because of Einstein's *local*-time emphasis, fell into disrepute. Absence of dynamical laws prevented Milne (despite his identifying the Lemaître-Hubble redshift 'constant' with the inverse of universe age) from persuading contemporaries of utility for Lorentz-group-founded cosmology. We resurrect Milne's cosmology by a Planck-scale slicing of spacetime that, in concert with *GN*'s source-bundle Fock space, accommodates Feynman-path [4] quantum-gravitational dynamics.

Absent slicing, Milne described the universe-encompassing forward-lightcone interior as 'continuously filled' with (unbounded) 3-dimensional metricized spacelike hyperboloids, the age of any hyperboloid being that of all its spacetime locations. Hyperboloid curvature is inversely proportional to hyperboloid age. Temporal tendency toward Euclid's ('ordinary' 3-space) geometry—a limit approached by Milne's geometry as age increases--classically endows

'age arrow' with local as well as global significance. Notice that, even without spacetime slicing, reproducibility of a 'measurement' (or of any local-history portion) is precluded by Milne's perpetually-ongoing flattening of 3-space.

A *QMU* sequence of exceptional 3-hyperboloids of age $N\delta$, with $N$ a positive integer and $\delta$ a universal Planck-scale age interval, [5] each house a fiber-bundle Fock-space ray. Milne relativity (*MR*) means that (at any age) two *QMU*'s related by a *global 7-SG* transformation are the *same* universe. Each ray (after the first) maintains this equivalence through invariance of the Feynman-path action that determines the ray from the (one-age-unit younger) predecessor ray.

The 7-*SG* group is a subgroup of the 12-parameter group of transformations of a unimodular 2×2 complex matrix by left or right multiplication by some (other) such matrix. Any left-group element commutes with any right element. (6-parameter left and right subgroups) The *QMU symmetry* group combines all right transformations (6-parameter subgroup) with left multiplication by *real diagonal* unimodular matrices (1-parameter subgroup). Energy generates the latter, momentum and angular momentum the former.

Although Milne's hyperbolic 3-space is noncompact, *QMU* houses, via a *periodic* Hilbert-space constraint, only a finite number of gravity sources. *MR*, via the vanishing of total momentum and angular momentum of a finite (although currently huge) universe, encompasses Mach's principle. Perpetuation of initially-zero total-universe energy sustains Milne's 'cosmological principle'.

*SS* comprises the interiors of a sequence of 4-dimensional *slice*s that share a common width in age. Excluded are the 3-dimensional hyperbolic manifolds that separate these slices. *SS* definition further excludes spacetime locations whose age is smaller than $N_0\delta$--the age at which the universe 'began'. Unrecognized by Milne, $N_0$ is a huge Mersenne prime that (along with 2, 3, 7 and 127 is foundational to *QMU*, although in this paper introduced without *raison d'être*, and associates both to 'inflation' and to 'universe size'. (The author expects, via Hilbert-space theory, future mathematical shrinkage of the current gap separating 'dimensionality'-ignoring discrete number theory and topology from dimensionality-accommodating continuous Lie-group theory and geometry. *QMU* invokes all the foregoing domains of mathematics. Fueling our anticipation has been the discovery, within the last century, of group contraction.)

The 'particle' concept of discrete and *stable* positive-energy 'clump', foundational to the S matrix and to the accompanying 'measurement' concept, lacks *a priori QMU* status. *QMU* foundations include *temporally-unstable* 'nonmaterial' (*not* 'particle') gravity sources with both negative and positive energies. [5] (*QMU total* energy, remember, vanishes.) The S-matrix, prerequisite to particle-physics meaning for 'measurement', is a 'dynamically-emergent' concept associated to *approximate* temporal stability of certain special positive-energy source 'clumps'. The accuracy within limited scale ranges for approximations such as that of the S matrix stems from $N_0$ hugeness. We shall employ the acronym *FAPPP* to characterize an approximation that, within some range of scales, is reliable 'for all practical physics purposes'.

*QMU* foundation eschews notions either of 'observer' or of 'measurement-result probability'. An observer-independent 'reality', consistent with an elsewhere-detailed complex-number-based Feynman-path quantum dynamics, will here be exposed. Within the 4-dimensional slice immediately following any global ray, 'settled' (real) classical fields are prescribed through *expectations* over this ray of self-adjoint field operators. These fields and derivatives thereof notably include locally-conserved (real) *energy-momentum* and *electric-charge* current densities—continuous local reality that manifests Planck's constant through 'microscopic' correlations between electric charge, energy, momentum, and spatial location. ('Micro' spacetime scale is huge compared to the Planck scale of $\delta$ while tiny compared to the 'macro' scale of $N_0\delta$ or to the still-larger redshift-manifested 'Lemaître-Hubble' cosmological scale set by universe age.) Such correlations enable, *FAPPP*, microscale meaning for 'particle' as well as ordinary-language macroscale meaning for 'object'. Strikingly absent from below-defined reality-associated self-adjoint field operators are any (zero-Dalembertian) quantum *radiation fields*.

Any Fock-space ray sums 'tensor' products of complex normed functions of 6 *source-bundle* coordinates that span a 3-dimensional compact 'fiber space' above each point of a 3-dimensional noncompact invariantly-metricized 'base space' which independently represents the Lorentz group. Schwinger's earlier use of the term 'source' [6] is related to ours but was applied to an operator rather than to a Fock-space vector and was not related to gravity. We below connect *our* meaning to the generation of both gravity and electromagnetism. The energy (regardless of sign) of any *QMU* source generates gravitation; charged sources additionally generate electromagnetism.

*QMU* sources, despite emulating in Hilbert-space status *nonrelativistic* Dirac particles at some common (global) time, are neither particles nor fields. *FAPPP*, at GUT spacetime scales--larger than Planck scale but not 'hugely' larger--another paper will attach 'elementary-particle' meaning to a *special* approximately-stable and approximately-localized *positive*-energy *single*-source ray that not only unitarily but also *irreducibly* represents the 7-*SG*--a 'unirrep'. This ray's expectation, of the source's self-adjoint 4-vector (lightlike although not positive) energy-momentum operator, approximates elementary-particle mass. Approximate *stability* of some ray (approximate age independence) we presume requires *positivity* for the ray's energy expectation.

*QMU* meaning for the adjective 'macroscopic' derives from the universe's spatial curvature at its 'beginning'—set by $N_0\delta$. A measure of 'universe size' is provided by the expectation of an operator whose eigenvalues are logarithms of positive integers that specify *total* numbers of sources. Such size would vanish at a universe *single*-source beginning but, regardless of the initial ray at age $N_0\delta$, *QMU* 's 3-space began with macroscopic (~km) 3-space (inverse) curvature--far above 'particle' (microscopic) scale and even further above the Planck scale set by the *SS* width $\delta$. [4] For microscopic scales down to GUT scale wherever stably-clumped positive-energy matter is accompanied by small gravitational potential, local Poincaré (flat 3-space) invariance applies *FAPPP*.

Although the (*zero*-age, 'universe-surrounding') *LSS* forward-lightcone boundary admits 'big-bang' association, the spacetime location of the lightcone vertex is meaningless. There is no Poincaré cosmological-symmetry group. Nevertheless, *individual* sources provide local unitary representation of conserved energy, momentum and angular momentum as well as of nonconserved chirality.

Locality and globality are linked by a 'local Lorentz frame' associated to each source location in the 6-dimensional product of fiber and base spaces. In the local frame of any such location the time component of the 4-vector spacetime displacement from Milne-lightcone vertex is equal to location age while the 3 spatial components all vanish. A source's fiber coordinates -- three 'Euler' angles—specify for the local frame a spatial orientation tied to the source's velocity direction and chirality.

Local-frame orientation is parallel-transportable along a base-space geodesic to any other location within the same hyperboloid. *Iff* a geodesic parallels a source's *velocity*, the fiber-space location of a source moving along this geodesic remains unchanged. Two of the 3 fiber angles coordinate a 'tangent space' of (local-frame) *source-velocity* directions. Source wave-function dependence on the remaining angle is elsewhere shown to define a 'chirality' that distinguishes 'bosonic' sources from 'fermionic' (while excluding either 'Higgs' or 'Majorana' sources).

Milne's use of the Lorentz group differs from that of Einstein and Poincaré--whose global-time-ignoring ('physics-appropriate') representation of *flat* 3-space and reversible local *particle*-time has been called 'special relativity'. Applied to an *individual* source, a 'Milne-Lorentz' boost shifts source location in hyperbolic (base) 3-space (at fixed age), as well as changing source orientation.

At those exceptional ages where a ray is defined, the universe comprises *fluctuating* sets of individual sources each endowed with spatial location, velocity direction, chirality, energy (of indefinite sign), momentum and angular momentum, as well as with (nonfluctuating) 'electric-charge number' and 'quark number' (3 times baryon number). All source attributes associate in Dirac sense [1] to (generally non-commuting) self-adjoint operators on the *GN* source fiber-bundle Hilbert space which unitarily represents the *7-SG*.

Any source-velocity magnitude equals *c*; there is no self-adjoint velocity-magnitude operator. Only the direction of source velocity is variable. Dirac attempted to represent Hilbert-space *sublightlike* motion through expectations over quantum-*fluctuating* lightlike-velocity direction— i.e., through 'zitterbewegung'-- [1] although failing because finite-dimensional Lorentz-group representations are not unitary. Via *GN*, we accomplish his objective. Fiber further allows Hilbert-space definition of a source ('reversible') local time that is independent of, while in Feynman paths dynamically consistent with, (irreversible) universe age. Conjugate to source (local) time in Dirac sense is source (local-frame) energy. A gravitational-potential operator will, three sections later, be defined by a sum of self-adjoint individual-source potentials each proportional to a source-energy operator defined two sections below.

**Source Fiber-Bundle Hilbert Space**

Source Hilbert-space vectors are functions of the coordinates of a 6-dimensional SL(2,c) manifold. Roughly speaking, three dimensions spatially locate a source, two specify its velocity direction and one its chirality. *GN* [(2)] defined a Hilbert space of differentiable normed complex functions of a unimodular 2×2 complex matrix ***a*** through *three* complex variables *s*, *y*, *z* (equivalent to six real variables) according to the following product of three unimodular 2×2 matrices, each of which coordinates the manifold of an abelian 2-parameter 7-*SG* subgroup:

$$\boldsymbol{a}(s, y, z) = exp(-\boldsymbol{\sigma}_3 s) \times exp(\boldsymbol{\sigma}_+ y) \times exp(\boldsymbol{\sigma}_- z). \tag{1}$$

Here $\boldsymbol{\sigma}_1$, $\boldsymbol{\sigma}_2$, $\boldsymbol{\sigma}_3$ is the standard (handed) set of Pauli hermitian traceless self-inverse 2×2 matrices (determinant −1), with $\boldsymbol{\sigma}_\pm \equiv \frac{1}{2}(\boldsymbol{\sigma}_1 \pm i\boldsymbol{\sigma}_2)$. The matrix $\boldsymbol{\sigma}_3$ is real diagonal while the real hermitian-conjugate matrix pair, $\boldsymbol{\sigma}_+$ and $\boldsymbol{\sigma}_-$, each has a unit off-diagonal element. Although the fiber-bundle nature of a source is better displayed by coordinates *alternative* (while equivalent) to *s*, *y*, *z*, we focus first on *GN*'s coordinate set as we proceed toward definition of classical-field reality.

We employ Dirac's shorthand [(1)] of denoting, by a *single* symbol, *both* a (real classical) coordinate *and* a self adjoint *operator* whose spectrum comprises the possible values of this coordinate. An example is the symbol *Re s*$^\sigma$--linearly related (we shall see) to the 'local time of Source $\sigma$'. The symbol $E^\sigma$ will denote Source-$\sigma$ local-frame energy. 'Canonically-conjugate' individual-source operators for local-time and local-frame-energy do not commute.

A discrete $\sigma$ *superscript* serves not only to designate an 'individual source' but to remind the reader of Dirac's quantum-classical dualism. Context will here distinguish self-adjoint operators from their classical counterparts.

The 6-dimensional 'volume element' (Haar measure),

$$d\boldsymbol{a} = ds\, dy\, dz, \tag{2}$$

is invariant under $\boldsymbol{a} \to \boldsymbol{a}^\Gamma \equiv \boldsymbol{a}\, \Gamma^{-1}$, with $\Gamma$ a 2×2 unimodular matrix representing a (Milne-sense) *right* Lorentz transformation of the coordinate ***a***. The measure (2) is also invariant under analogous left transformation. The volume-element symbol $d\xi$, with $\xi$ complex means $d\, Re\, \xi \times d\, Im\, \xi$.

Any universe ray has age $N\delta$ with $N$ a (positive) integer. An $N$-Hilbert-space (*single*) $\sigma$-labeled 'gravity-source' vector is a complex differentiable and periodic function $\psi^N(\boldsymbol{a}^\sigma)$ with invariant (finite) norm,

$$\int d\boldsymbol{a}^\sigma \left| \psi^N(\boldsymbol{a}^\sigma) \right|^2 . \tag{3}$$

Invariant *periodicity* is addressed in an Appendix.

The Fock-space $N$ ray $\Psi^N$ is a sum of normed functions of *differing* numbers, *n*, of the source coordinates $\boldsymbol{a}^\sigma$:

$$\Psi^N = \Sigma_n\, \psi^N_n (\boldsymbol{a}^1 \dots \boldsymbol{a}^n), \tag{4}$$

the number of sources *n* running between 1 and, for Feynman-path reasons exposed elsewhere, a maximum determined by $N-N_0$. 'Vacuum'—i.e., $n = 0$--is absent from *QMU* Fock space.

The norm-defining volume element for functions of *n* $\sigma$-labeled source coordinates is a product of single-source volume elements each having the form (2). With a notation $\{O\}_N$ denoting Ray-$N$ expectation of a self-adjoint operator *O*, the non-negative bounded dimensionless expectation, $\{ln\ n\}_N$, defines a 'universe size' that, at any human-history age, is huge even though finite.

A superscript $\sigma$, as in (3), associated to a ray-labeling age integer, $N \geq N_0$, designates a 'source' within some set of $n$ sources, as in (4). Understanding the *meaning* of $\sigma$ to include some value of $n$, the present paper henceforth will dispense with the $n$ symbol. We shall use the symbol $\tau$ to designate age (global time), whether a discrete ray age—$\tau = N\delta$--or a continuous *SS*-interior age.

A 'Milne-Lorentz' SL(2,c) transformation specified by the 2×2 complex unimodular *right*-acting matrix $\Gamma$ is *unitarily* Source-$\sigma$ Hilbert-space represented by

$$\Psi^N(\boldsymbol{a}^\sigma) \rightarrow \Psi^N(\boldsymbol{a}^\sigma \Gamma^{-1}). \tag{5}$$

Calculation shows $\boldsymbol{a}\Gamma^{-1}$ (here omitting the source-label $\sigma$) to be equivalent to

$$z^\Gamma = (\Gamma_{22}z - \Gamma_{21})/(\Gamma_{11} - \Gamma_{12}z), \tag{6}$$

$$y^\Gamma = (\Gamma_{11} - \Gamma_{12}z)[(\Gamma_{11} - \Gamma_{12}z)y - \Gamma_{12}], \tag{7}$$

$$s^\Gamma = s + ln\ (\Gamma_{11} - \Gamma_{12}z). \tag{8}$$

Notice that the 2-dimensional volume element $ds$ within the Haar measure is *separately* invariant (implying invariance also of the 4-dimensional volume element $dy\ dz$). This invariant factorizability of measure dovetails with source energy and chirality definition and the Appendix-prescribed periodicity constraint on source Hilbert space.

The fiber-bundle character of a source is exhibited by a factorization of the 6-dimensional single-source manifold which is *alternative* to that of *GN*. The 2×2 unimodular matrix $\boldsymbol{a}$ equals the product $\boldsymbol{uh}$ of a 3-parameter unitary unimodular matrix $\boldsymbol{u}$ and a 3-parameter *positive* hermitian unimodular matrix $\boldsymbol{h}$. The matrix $\boldsymbol{u}$ may be parameterized by 3 real ('Euler') angles, $0 < \varphi' < 4\pi$, $0 < \theta < \pi$, $0 < \varphi < 2\pi$, such that

$$\boldsymbol{u} = exp\ (i\sigma_3\ \varphi'/2)\ exp\ (i\sigma_1\theta/2) \times exp\ (i\sigma_3\ \varphi/2), \tag{9}$$

and the matrix $\boldsymbol{h}$ by a real 3-vector $\boldsymbol{\beta}$ such that

$$\boldsymbol{h} = exp\ (-\boldsymbol{\sigma}\cdot\boldsymbol{\beta}/2), \tag{10}$$

with the ·symbol in (10) denoting the rotationally-invariant inner product of two 3-vectors.

Calculation shows that in source local frame, where $\boldsymbol{\beta}$ vanishes, $Im\ s = -\varphi'/2$, $Re\ s = -\frac{1}{2}\ ln\ [1+|z|^2]$, $y = -z^*/(1+|z|^2)$, while $z = i\ e^{i\varphi}\ tan\ \theta/2$. Generally, as elaborated below, $\boldsymbol{\beta}$ specifies source spatial (base-space) location while $\boldsymbol{u}$ specifies local-frame source orientation—i.e., location in fiber space. The 2-dimensional fiber subspace coordinated by $\theta, \varphi$, we refer to as the base space's 'tangent' subspace or, alternatively, as source local-frame 'velocity space'. The circle doubly spanned by $\varphi'$ might be called 'source-chirality' space.

The Haar measure (2) equals a product of two separately-invariant 'volume elements': fiber volume, $d\mathbf{u} = d\varphi' \sin\theta\, d\theta\, d\varphi$, and base volume, $d\boldsymbol{\beta} = \sinh^2\beta\, d\beta\, d\Omega$, where $\boldsymbol{\beta} = \beta\mathbf{n}$ with $\beta > 0$ and $\mathbf{n}$ a unit 3-vector. The symbol $d\Omega$ denotes an infinitesimal solid angle at the direction $\mathbf{n}$. The 'base' 3-space coordinated by $\boldsymbol{\beta}$ enjoys an invariant 'dimensionful' (age-related) metric while the compact fiber 3-space coordinated by $\mathbf{u}$ does not. (Lorentz transformation of location in fiber space is 'entangled' with location in base space although the converse is not true. The fiber's Haar measure is a product of two *separately 7-SG invariant*—velocity-direction and chirality--volume elements.)

**Self-Adjoint Single-Source Operators**

A base-space *geodesic* is spanned by the 1-parameter group of Haar-measure-preserving *left* transformations, $\mathbf{a}^\sigma \to [\exp(-\boldsymbol{\sigma}_3 \tilde{a})]\, \mathbf{a}^\sigma$ with $\tilde{a}$ real and positive. Such a *left* transformation commutes with all Milne-Lorentz (*right*) transformations)--increasing $Re\, s^\sigma$ by $\tilde{a}$ at fixed $Im\, s^\sigma$, $y^\sigma$ and $z^\sigma$. Because a source is thereby spatially displaced in its velocity direction, [5] a '*source--time*' increase becomes defined.

A local-frame source-$\sigma$ Hilbert-space *energy* operator emerges from Fourier transformation of Source-$\sigma$ wave-function dependence on $Re\, s^\sigma$. A self-adjoint operator $E^{N,\sigma}$, representing the Source-$\sigma$ local-frame energy, is definable in the 'spacetime' $(s, y, z)$ basis by

$$E^{N,\sigma} \equiv i(\hbar/N\delta)\, \partial/\partial\, 2Re\, s^\sigma, \tag{11}$$

the partial derivative being at fixed $Im\, s^\sigma$, $y^\sigma$, and $z^\sigma$ as well as at fixed $a^\rho$ for $\rho \neq \sigma$. The indefinite-sign-spectrum of $E^{N,\sigma}$ is 7-SG invariant. In Dirac sense $E^{N,\sigma}$ is *conjugate* to the self-adjoint source-time operator $t^{N,\sigma} \equiv N\delta[2Re\, s^\sigma + \ln(1+|z^\sigma|^2)]$, that vanishes in the source's local frame and whose increase another paper will coordinate with age increase along a Feynman path.

A 4-vector with Lorentz index $\mu = 0, 1, 2, 3$ is equivalent to a hermitian 2×2 matrix $\mathbf{\mathfrak{h}}$ through the formula $\frac{1}{2}tr\, \sigma_\mu\, \mathbf{\mathfrak{h}}$. By $\sigma_0$ is meant a unit matrix. Omitting for a few paragraphs the source label, two (commuting, classical or operator) positive ($tr\, \mathbf{\mathfrak{h}} > 0$) 4-vectors, for $\tau = N\delta$,

$$\mathbf{x}^\tau \equiv \tau\, \mathbf{a}^\dagger \mathbf{a}\ [= \tau\, \mathbf{h}^2 = \tau\, \exp(-\boldsymbol{\sigma}\cdot\boldsymbol{\beta})], \tag{12}$$

$$= \tau\, [\sigma_0 \cosh\beta - \boldsymbol{\sigma}\cdot\mathbf{n}\, \sinh\beta], \quad (\boldsymbol{\beta} = \beta\mathbf{n},\, \mathbf{n}\cdot\mathbf{n} = 1,\, \beta \geq 0), \tag{12'}$$

and

$$\mathbf{v} \equiv \mathbf{a}^\dagger\, (\sigma_0 - \sigma_3)\, \mathbf{a}, \tag{13}$$

satisfying three Milne-Lorentz-covariant inner-product (homogeneous-quadratic) constraints, are equivalent to either quantum or classical meanings for the coordinate quintet $Re\, s,\, y,\, z$, of spatial-location and velocity. I.e., this quintet is a set of 5 commuting self-adjoint operators equivalent to the location-velocity 4-vector operator pair $\mathbf{x}^\tau,\, \mathbf{v}$. The (missing, sixth) coordinate-operator, $Im\, s$, associates to source chirality (to fermionic-source helicity).

The trio of 4-vector constraints, built into (12) and (13), are $\mathbf{x}^\tau \bullet \mathbf{x}^\tau = \tau^2$, $\mathbf{x}^\tau \bullet \mathbf{v} = \tau$, $\mathbf{v} \bullet \mathbf{v} = 0$, where the symbol $\bullet$ denotes a Lorentz inner product of 4-vectors. These constraints (applying *both* to real coordinates *and* to commuting self-adjoint operators) complicate expression through the 4-vector pair of the 5-dimensional (chirality-ignoring) Haar-measure.

The positive-timelike 4-vector $x^\tau$ specifies a source's spacetime location with respect to Milne-lightcone vertex while the positive-lightlike 4-vector $v$ specifies the source's velocity (in $c$ units). In source local frame (i.e., locating Source-$\sigma$ at base-space 'origin') the location 4-vector components, which generally are $\tau$ (*cosh β*, $n$ *sinh β*), become ($\tau$, 0, 0, 0) while the ('tangent-space') velocity-4-vector components become (1, *sin θ cos φ*, *sin θ sin φ*, *cos θ*).

Dependence on the operator *Re s* by both $x^\tau$ and $v$ [revealed by Formulas (12) and (13)], means from (11) that neither the source-location operator nor the source-velocity operator commutes with the (invariant) source local-frame-energy operator (11). We nevertheless may define a self-adjoint indefinite-sign single-source energy-momentum lightlike 4-vector operator,

$$p^{N,\sigma} \equiv \tfrac{1}{2}(E^{N,\sigma}v^\sigma + v^\sigma E^{N,\sigma}), \tag{14}$$

whose (real) 4-vector *expectation*, $\{p^{N,\sigma}\}_N$, need be neither positive lightlike nor timelike. A source's 'mass squared' –defined by $\{p^{N,\sigma}\}_N \bullet \{p^{N,\sigma}\}_N$ –generally may be either positive or negative, as also may be its energy. Only when $\{p^{N,\sigma}\}_N$ and $\{p^{N,\sigma}\}_N \bullet \{p^{N,\sigma}\}_N$ are *both* non-negative may a single-source wave function associate to 'matter'.

The 3-vector components of the dimensionful 4-vector (14) are *different* from the (Appendix-addressed) 3-vector dimensionful generators of the Milne-Lorentz group (components of a 6-vector). *Special QMU* single-source wave functions with approximately equal 'lab-frame' expectations for the two *different* momentum operators are particle candidates.

Expectation 4-vectors of the operator (14) provide (approximate) *QMU* contact with the Poincaré-group-representing (momentum-basis) S matrix. Although *QMU* 'classical reality' (following section), fails to single out individual sources, we expect the ('standard') S-matrix meaning for 'massive elementary particle' to connect with certain real-mass, positive-energy single-source wave functions that are approximately *stable*—i.e., approximately $N$ independent.

As proved true for Dirac's 'relativistic electron,'[1] the *QMU ratio* of *stable*-source momentum and energy expectations we anticipate will equal (approximately) the expectation of source *velocity*. ('Dark matter' we associate to positive-energy gravity sources so unstable as to be inaccessible to particle physics—i.e., to be S-matrix *indescribable*.)

**Classical-Field Spacetime Reality**

We postulate a retarded Source-$\sigma$-generated ('Lienard-Wiechert'—*LW*) divergenceless gravitational potential that is a self-adjoint *field* operator--a function of the Source-$\sigma$ operators (11), (12) and (13) which depends *also* on a classical 'sink' location in spacetime. Our 'Newton-*LW*-Dirac-Milne' gravitational-potential operator transforms as a symmetric second-rank Lorentz tensor of zero invariant trace. A sum over *all* Ray-$N$ sources then defines a global operator whose expectation over that ray is *the* (classical) gravitational potential throughout the immediately-following *SS*. This potential's Dalembertian (when divided by $G$) prescribes within the slice the conserved current density of energy-momentum.

Our postulate for the retarded gravitational-potential operator $\Phi^{N,\sigma}_{\mu\upsilon}(x)$, associated to Source-$\sigma$ of age $N\delta$ and to sink spacetime location $x$ (the 'field-point'), is

$$\Phi^{N,\sigma}{}_{\mu\upsilon}(\mathbf{x}) = G\, c^{-5}\, [E^{N,\sigma}\, V^{N,\sigma}{}_{\mu\upsilon}(\mathbf{x}) + V^{N,\sigma}{}_{\mu\upsilon}(\mathbf{x})\, E^{N,\sigma}], \tag{15}$$

where

$$V^{N,\sigma}{}_{\mu\upsilon}(\mathbf{x}) \equiv \Theta^{N}{}_{ret}(\mathbf{x}, \mathbf{a}^{\sigma})\, v^{\sigma}{}_{\mu}\, v^{\sigma}{}_{\upsilon} / v^{\sigma} \bullet (\mathbf{x} - \mathbf{x}^{N\delta,\sigma}), \tag{16}$$

with the retardation step function $\Theta^{N}{}_{ret}(\mathbf{x}, \mathbf{a}^{\sigma})$ to be defined two paragraphs below. We suppose the field-point location $\mathbf{x}$ to lie within the $N$-ray immediate future, where

$$(N\delta)^2 < \mathbf{x} \bullet \mathbf{x} < [(N+1)\delta]^2. \tag{17}$$

The $\mathbf{x}$ dependence of our potential is seen to reside in the invariant *LW*-denominator operator, $v^{\sigma} \bullet (\mathbf{x} - \mathbf{x}^{N\delta,\sigma})$. Because $v^{\sigma} \bullet v^{\sigma} = 0$, this *classical* denominator-polynomial has the same value at *all* 'potential-source' spacetime locations (not only those of age $N\delta$) along the lightlike trajectory with Source-$\sigma$ velocity that passes through $\mathbf{x}^{N\delta,\sigma}$. If the $\mathbf{a}^{\sigma}$ trajectory *intersects* the $\mathbf{x}$ backward lightcone we choose, in *classical* language, to call that intersection's location *the* spacetime location of the 'retarded source' for $\Phi^{N,\sigma}{}_{\mu\upsilon}(\mathbf{x})$. Sources 'classically located' in the 'distant past' of $\mathbf{x}$ then associate to Age-$N\delta$ ('near past') *Hilbert-space* sources whose *spatial* location is far from that of the field point $\mathbf{x}$.

The symbol $\Theta^{N}{}_{ret}(\mathbf{x}, \mathbf{a}^{\sigma})$ in (16) denotes a step function equal to 1 *iff* the $\mathbf{a}^{\sigma}$ trajectory (passing with velocity $v^{\sigma}$ through the spacetime location $\mathbf{x}^{N\delta,\sigma}$) intersects the $\mathbf{x}$ *backward* lightcone. Otherwise $\Theta^{N}{}_{ret}(\mathbf{x}, \mathbf{a}^{\sigma})$ vanishes. (*Any* lightlike trajectory not located *on* the $\mathbf{x}$ lightcone intersects the $\mathbf{x}$ forward-backward lightcone exactly *once*.) Summed over all sources, the Ray-$N$ expectation of (15) prescribes the classical gravitational tensor potential $\Phi^{N}{}_{\mu\upsilon}(\mathbf{x})$ within the $N$ immediate future.

Because the source-velocity 4-vectors $v^{\sigma}$ are lightlike, the Lorentz-divergence of $\Phi^{N}{}_{\mu\upsilon}(\mathbf{x})$ vanishes. When divided by $G$, the Dalembertian of $\Phi^{N}{}_{\mu\upsilon}(\mathbf{x})$ prescribes within the $N$ immediate future (without Heisenberg uncertainty) the 4-vector current density of conserved energy-momentum—'gravitational reality'.

Although a symbol $\Phi_{\mu\upsilon}(\mathbf{x})$, *without* superscript $N$, conveniently designates the gravitational retarded potential *almost everywhere* in Milne spacetime, exclusion must be remembered of ray ages where $\tau = N\delta$. Classical fields are not defined *on* the exceptional hyperboloids that house rays. Differential equations for (classical) gravitational fields and energy-momentum current densities, meaningful *inside* any *SS*, only *approximately* extrapolate classical fields from one slice to the next. (In humanity-occupied universe such approximation often suffices *FAPPP*.) Feynman paths determine universe evolution *quantum-mechanically*--via action-specified *phases* of complex numbers.

Paralleling Formulas (15) and (16) we define (within the *SS* following Ray $N$) a divergenceless *electromagnetic* Source-$\sigma$ retarded vector potential operator whose dimensionality is that of action divided by time:

$$A^{N,\sigma}(\mathbf{x}) \equiv \Theta_{ret}(\mathbf{x}, \mathbf{a}^{\sigma})\, Q^{\sigma}\, v^{\sigma} / v^{\sigma} \bullet (\mathbf{x} - \mathbf{x}^{N\delta,\sigma}). \tag{18}$$

Here the symbol $Q^{\sigma}$ designates source electric charge. The Ray-$N$ expectation of $A^{N,\sigma}(\mathbf{x})$, summed over all sources, then yields the (classical) divergenceless electromagnetic vector potential *A*(*x*) within Ray *N*'s immediate future. (Represented by this expectation, when applied to the *FAPPP*-stable source wave function of a single charged electron, is an effect QED calls 'charge renormalization'.) With an appropriate factor the Dalembertian of *A*(*x*) specifies 'electromagnetic reality' as a locally conserved electric-charge current density. Maxwell's (classical) equations for electric and magnetic fields apply *inside* each spacetime slice.

**Conclusion**

This paper has quantized Milne's cosmology so as to represent gravity while maintaining Dirac's principles as well as that of Mach. A 7-parameter symmetry group (with a 6-parameter Lorentz subgroup) is accompanied by 7 (Noether) conserved self-adjoint group-generating operators— momentum, angular momentum and energy—whose expectations separately aggregate to zero for the universe as a whole. Classical reality resides in electromagnetic and gravitational fields within invariant spacetime slices of Planck-scale width. Each universe ray is separated from its successor by such a slice. Any ray (after the first) is determined from its predecessor by the actions of Feynman paths that traverse the preceding slice. Gravitational and electromagnetic line-integral path action, determined by the potentials here defined, are specified in a separate paper. Another paper in preparation proposes at path 'branchings' an *event action* that imitates GUT quantum field theory while maintaining gravity and Dirac-Feynman principles. Avoided are (Standard-Model) arbitrary parameters to specify scales, elementary-particle masses and fermion-generation mixing.

**Appendix: Momentum Operators; Dimensionality; Hilbert-Space Periodicity**

A trio of mutually-*noncommuting* self-adjoint source-momentum operators, components of a Lorentz 6-vector (*not* a 4-vector), transform under rotations as a 3-vector whose components commute with Formula (11)'s source-energy operator. The latter generates a Hilbert-space unitarily-represented *single*-parameter Lie group comprising displacements in *source* time at fixed universe age—3-space velocity-directed displacements along some base-space geodesic.

Formula (A.1) below is a self-adjoint superposition of (partial) first derivatives, *different* from (11), that represents the component of source *momentum* in an *externally-fixed* direction. The standard notation for Pauli matrices suggests calling the latter (*not* the source's-velocity direction) the '3-direction'; the 3-direction within hyperboloids may arbitrarily be assigned. Formulas (6), (7) and (8) prescribe the 3-direction source-momentum operator to be

$$P_3^{N,\sigma} \equiv i(\hbar/c_N\delta)\,[\partial/\partial 2Re\,s^\sigma + \partial/\partial ln|y^\sigma| - \partial/\partial ln|z^\sigma|]. \tag{A.1}$$

Held fixed in (A.1) are the coordinates $arg\,y^\sigma$, $arg\,z^\sigma$ and $2Im\,s^\sigma$. Partial derivatives with respect to the latter coordinate trio appear in a representation, paralleling (A.1), of 3-direction source *angular* momentum, $J_3^{\,\sigma}$. Four companion (by rotation) operators $P_1^{N,\sigma}$, $P_2^{N,\sigma}$, $J_1^{\sigma}$, $J_2^{\sigma}$ complete a 6-generator nonabelian Lorentz-group algebra.

Distinction is useful between 'dimensionful' and 'dimensionless' generators. 'Dimensionality' relates both to the dilation group and to the nonabelian Lorentz-group algebra.

In an infinite-age limit, the Milne-Lorentz group 'contracts' to the Euclidean group, whose spatial-displacement generators familiarly enjoy 'momentum' interpretation.

Units may be chosen such that $\hbar = c = 1$. Energy and momentum then have the same dimensionality, as do time and spatial displacement, while the latter pair's (shared) dimensionality is the *negative* of that of the former. Any self-adjoint *QMU* operator enjoys some dimensionality $q$, where $q$ is an integer. Action, angular momentum, velocity and electric charge are all 'dimensionless'—i.e., these operators and their expectations have $q = 0$.

So are *products* of energy and time displacement, as well as products of momentum and spatial displacement. ['Opposite values of $q$ are generally carried by the two members of any Dirac-sense 'canonically-conjugate' pair of self-adjoint operators.] The momentum generators of the Milne-Lorentz group are dimensionful (because the slice width $\delta$ has $q = 1$), in contrast to the angular-momentum generators--which are dimensionless.

Because Newton's gravitational constant has $q = 2$, the retarded single-source gravitational-potential self-adjoint operator (15), is dimensionless, as is the expectation of the corresponding sum over all sources—i.e., the classical gravitational potential, $\Phi_{\mu\upsilon}(x)$.

Although Milne's hyperbolic classical 3-space is unbounded, we 'compactify' *QMU* source Hilbert space by requiring

$$\psi^N[a^\sigma] = \psi^N[exp(-\tfrac{1}{2}\sigma_3 \Delta)a^\sigma]. \tag{A.1}$$

The foregoing periodicity in $2Re\, s^\sigma$, with a (dimensionless) period $\Delta$ that is *common* to all $\sigma$, we accompany by similar periodicity in all $ln|y^\sigma/z^\sigma|$. *Iff $\Delta$ is large*, the universe is 'large on Lemaître-Hubble scale'. The spectra of both source (local-frame) energy and source momentum are thereby universally discretized in units of $\hbar/\Delta N\delta$. Compactification allows an initial condition at $N = N_0$ to specify zero total universe energy, momentum and angular momentum--a condition perpetuated by the Feynman-path action that propagates each ray to its successor.

**Acknowledgements**